\documentclass[
a4paper,       
reprint,           
superscriptaddress,
amsmath,           
amssymb,           
prd,               
notitlepage,       
longbibliography,  
floatfix,          
twocolumn,
nofootinbib,          
]{revtex4-1}

\usepackage{subfigure}
\usepackage{caption}
\usepackage{ulem}
\usepackage{graphicx}
\usepackage{dcolumn}
\usepackage{bm}
\usepackage{tensor}     
\usepackage[
colorlinks=true,        
citecolor=blue,         
linkcolor=blue,         
urlcolor=blue           
]{hyperref}             
\usepackage{comment}

\def\d{\mathrm{d}}

\begin{document}

\preprint{APS/123-QED}
\title{Nonlinear dynamics of oscillons and transients during preheating after single field inflation}

\author{Tianyu Jia}
\email{jfree1999@163.com }
\affiliation{Center for Gravitation and Cosmology, College of Physical Science and Technology, Yangzhou University, Yangzhou 225009, China}

\author{Yu Sang}
\email{sangyu@yzu.edu.cn (corresponding author)}
\affiliation{Center for Gravitation and Cosmology, College of Physical Science and Technology, Yangzhou University, Yangzhou 225009, China}

\author{Xue Zhang}
\email{zhangxue@yzu.edu.cn}
\affiliation{Center for Gravitation and Cosmology, College of Physical Science and Technology, Yangzhou University, Yangzhou 225009, China}

\date{\today}

\begin{abstract}
In the single-field model, the preheating process occurs through self-resonance of inflaton field.
We study the nonlinear structures generated during preheating in the $\alpha$-attractor models and monodromy models. 
The potentials have a power law form $\propto\left|\phi\right|^{2n}$ near the origin and a flat region away from bottom, which are consistent with current cosmological observations. 
The Floquet analysis shows that potential parameters in monodromy model have a significant influence on the region of resonance bands. 
Besides we investigate the formation of nonlinear structures, the equation of state and the energy transfer through the (3+1) dimensional lattice simulation. We find that all the models exhibit similar nonlinear dynamics. 
Furthermore, our analysis reveals that the behavior of the inflation model at small field values, which is governed by the parameter $n$, exerts a more substantial influence on the nonlinear dynamics. The behavior at large field values, controlled by the parameter $q$, has a negligible impact.

\end{abstract}

\maketitle

\section{Introduction}
The single field inflation driven by a power-law potential is ruled out by the measurements of cosmic microwave background (CMB) anisotropies \cite{Planck:2018,Planck:2018constraitn}. The observational constraints favor a broad class of potentials characterized by a power-law shape at the bottom and plateaus in region away from the minimum \cite{Cai:2015soa,Cook:2015vqa,Kabir:2016kdh,Martin:2014nya,German:2023yer}. These class of models has interesting nonlinear dynamics in the post-inflationary evolution \cite{Hindmarsh,Enqvist:2002Q-ball,Coleman:1985kiQ-ball,Frolov:2008code-bubble,Amin2018kkgdomain-wall,Karam:2023PBH}. After the end of inflation, the inflaton field oscillates around the minimum of its potential \cite{Guth,Mukhanov:1981xt,Starobinsky:1982ee,Linde:1983chaotic_inflation,Linde:1990flp}. 
The energy transfer from the inflaton field to standard model particles occurs through the oscillating decay process, leading to the subsequent thermalization of the universe and the onset of a radiation-dominated era \cite{Albrecht:1982mp}.
In traditional preheating scenarios, the fluctuations of the coupling matter field undergo exponential amplification through parametric resonance \cite{Dolgov:1989us,Traschen:1990sw,Kofman:1997yn,Shtanov:1994ce}. However in the models with plateaus-like potential, self-resonance occurs and leads to the amplification of the inflaton field itself \cite{Amin:2012}.

Oscillon is a kind of long-lived, localized excitations of nonlinear scalar fields \cite{Kasuya:2002adiabatic,Amin:2010xe,Amin:2010dc,Shafi:2024jig,Piani:2023aof,Zhang:2020bec,Zhang:2020ntm,Zhang:2024bjo}. If the inflaton potential has a quadratic bottom and flattens away from the minimum, oscillons are copiously produced as a consequence of self-resonance \cite{Alp-oscillon,Mahbub:2023faw,Sang:2020kpd}. 
A possible matter-dominated phase exists if oscillons are fully formed in the post-inflationary stage, which has been proved by simulating the equation of state in some models \cite{P.R.L.2017,Lozanov:2018hjm,Liu:2017huacospy.potential}.
Gravitational waves are generated when the oscillons are forming, because of the time-evolving and inhomogeneous energy density \cite{PhysRevD.2019,Liu:2018rrt,WangqingyangGW,Zhou-shuang-yongGW,Sang:2019ndv,Hiramatsu:2020GW,Li:2020qnk,Amin:2018xfe,Antusch:2016con,Fu:2019qqe,Jin:2020tmm,Adshead:2023nhk,Adshead:2024ykw,ElBourakadi:2021nyb}.

Another kind of nonlinear structure was found to be formed during preheating after the $\alpha$-attractor models which are consistent with the CMB observation and attract lot of interest of the community \cite{Ueno:2016constraintalphamodel,ElBourakadi:2022lqf,Eshaghi:2016kne}. The equation of state and duration to radiation domination were calculated in the $\alpha$-attractor T model for both quadratic minima and nonquadratic minima cases \cite{P.R.L.2017,Lozanov:2018hjm}. If the minimum shallower than quadratic, e.g., $ V \propto |\phi|^{2n}$ by $n > 1$, the inflaton condensates fragment into transients, which are oscillon-like localised spherical objects but much less stable. The equation of state is similar to radiation ($w \rightarrow 1/3$) in the case of transients formation, as oppose to the oscillon-dominated phase ($w \rightarrow 0$).

The small amplitude analysis is employed to examine the dynamics of the field and identify any peculiar field configurations \cite{Fodor:2006zs,Farhi:2007wj,PhysRevD.78.025003,quantum_radiation-Hertzberg:2010yz}. 
For the case of oscillons, there exists an oscillating solution with a stable configuration \cite{a-efragment,Flat-top.oscillons,Amin:2013k-oscillons,Manton:2023mdr,Cyncynates:2021rtf}. 
While in the case of transient, the resonance term disappears and the dynamic behavior of the field will no longer be oscillatory, which means that the small amplitude analysis method is not applicable to the transient solution at a higher order \cite{ Lozanov:2018hjm}.  And an effective method for studying transient dynamics is lattice simulation.

In this paper we consider a class of observationally favored inflation models, taking $\alpha$-attractor T model, E model and monodromy-like models as examples. We will emphasize several key questions within these models, including whether symmetric potential in T model and the asymmetric potential in E model lead to distinct nonlinear dynamics, how different parameters in monodromy models affect the the formation of oscillons and trasients, the equation of state during the evolution.

This paper is organized as follows.
In Sec. \ref{Sec:models} we introduce the inflation models used in this work.
We give a linear Floquet analysis in Sec. \ref{Sec:Floquet}. The lattice simulation method is introduced in Sec. \ref{Sec:Lattice method}, and the simulation results are given in Sec. \ref{sec:alpha attractor model} for $\alpha$-attractor models and Sec. \ref{sec:axion monodromy models} for monodromy models. In Sec. \ref{sec:conclusion} we summarizes our work along with future prospects. We adopt natural units with $G=\hbar=c=1$ and the reduce Planck mass $m_{\rm pl}=1/\sqrt{8\pi G}$. The metric signature is chosen as $(+ - - -)$, and we employ the Friedmann-Robertson-Walker metric in the form of $\d s^{2}=\d t^{2}-a^{2} \d x^{2}$, where $a$ represents the scale factor.

\begin{figure*}[!htbp]
    \centering
    \subfigure{
    \begin{minipage}[b]{.22\textwidth}
    \centering
    \includegraphics[width=\linewidth]{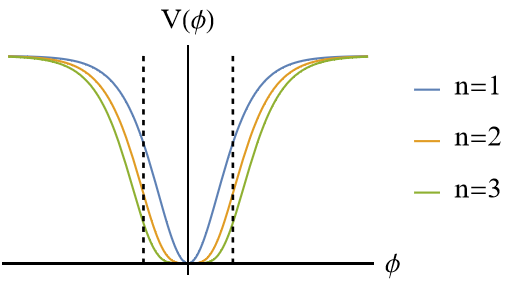}
    \caption*{(a) T potential}
    \end{minipage}
    }
    \centering
    \subfigure{
    \begin{minipage}[b]{.22\textwidth}
    \centering
    \includegraphics[width=\linewidth]{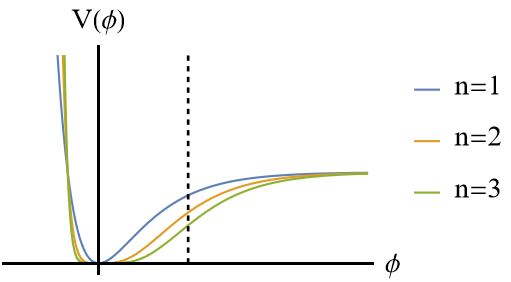}
    \caption*{(b) E potential}
    \end{minipage}
    }
    \centering
    \subfigure{
    \begin{minipage}[b]{.22\textwidth}
    \centering
    \includegraphics[width=\linewidth]{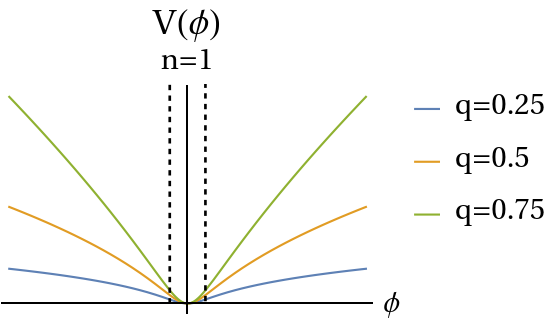}
    \caption*{(c) monodromy potential}
    \end{minipage}
    }
        \centering
    \subfigure{
    \begin{minipage}[b]{.22\textwidth}
    \centering
    \includegraphics[width=\linewidth]{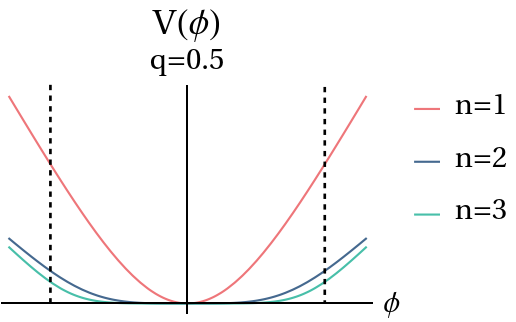}
    \caption*{(d) monodromy potential}
    \end{minipage}
    }
    \caption{The potentials of $\alpha$-attractor T model, E model and monodromy model used in this work. The dotted line represents the inflection point of the curve. Panel (c) shows the large field region and  Panel (d) shows the small field region of monodromy model.}
    \label{fig:potential}
\end{figure*}
\par

\section{models}\label{Sec:models}

We consider the single field inflation model characterized by minimal coupling with gravity. The action is expressed as follows
\begin{equation}\label{eq:action}
\begin{aligned}
    S=\int \mathrm{d}^{4}x\sqrt{-g}\left[-\frac{m_{\rm pl}^{2}R}{2}+\frac{1}{2}\partial_{\mu}\phi\partial^{\mu}\phi-V\left(\phi\right)\right],
\end{aligned}
\end{equation}
where $R$ represents the Ricci scalar, $\phi$ denotes the inflaton field. We consider the Friedmann-Robertson-Walker (FRW) metric here. The equation of motion for the scalar field and the Friedmann equation have the form
\begin{equation}
\begin{gathered}
    \ddot{\phi}+3H\dot{\phi}-\frac{\nabla^{2}\phi}{a^{2}}+\frac{\mathrm{d}V\left(\phi\right)}{\mathrm{d}\phi}=0,\label{eq:motion}
\end{gathered}
\end{equation}
\begin{equation}
\begin{gathered}
    H^{2}=\frac{1}{3m_{\rm pl}^{2}}\left(\frac{\dot{\phi}^{2}}{2}+\frac{\left(\nabla\phi\right)^{2}}{2a^{2}}+V\left(\phi\right)\right). \label{eq:Friedmann}
\end{gathered}
\end{equation}

In this work, we consider a class of inflation models that is consistent with the constraint of CMB observations  \cite{Planck:2018constraitn}. 
The inflaton potential exhibits a plateau at large field values $\left(\phi\gg M\right)$ and a power law form $\propto |\phi|^{2n}$  near small field values ($\phi\ll M$). We chose three distinct  parameterized models:
\begin{itemize}
    \item the $\alpha$-attractor T models \cite{Carrasco:tmodel,Kallosh:conformal.inflation,Galante:attractor}
\begin{equation}
    V\left(\phi\right)=\Lambda^{4} \tanh^{2n}\left(\frac{\left|\phi\right|}{M}\right),
\end{equation}
    \item the $\alpha$-attractor E models \cite{Carrasco:tmodel,Kallosh:conformal.inflation,Galante:attractor}
\begin{equation}
    V\left(\phi\right)=\Lambda^{4}\left| 1-\exp\left(-\frac{2\phi}{M}\right)\right|^{2n},
\end{equation}
    \item monodromy type potentials \cite{McAllister:Axion.inflaiton,McAllister:2014mpaaxion.model,Silverstein:2008sgaxion.model}
\begin{equation}\label{eq:axion model}
    V\left(\phi\right)=\Lambda^{4}\left[\left(1+\left|\frac{\phi}{M}\right|^{2n}\right)^{\frac{q}{2n}}-1\right].
\end{equation}
\end{itemize}

The potentials of $\alpha$-attractor T model, E model and monodromy model are shown in Fig. \ref{fig:potential}. The dotted line is the inflection point of the curve. In all models, the potential has a power-law bottom, and exhibits different asymptotic behavior far away from it. The parameter $M$ is related to the inflection point of the potential. In this work, we set the value of parameter $M=0.01m_{\rm pl}$, ensuring that all of them satisfy the observational boundary requirements $\left(M\ll10~m_{\rm pl}\right)$ and successfully implement the slow-roll process. Parameter $n$ determines its behavior near the bottom in all models. The energy scale parameter $\Lambda$ is eliminated using the amplitude of the scalar perturbations and spectral tilt from the CMB observations.

Both the T models and monodromy models are symmetric, while the E models are asymmetric. In the T models, the potential tends to a constant value at infinity, whereas in E models it reaches an extreme value on one side only. 
In addition, in monodromy models, there is an extra parameter $q$ that can be utilized to suppress the upward trend of the curve near the large field values. It controls the asymptotic behavior of the potential in infinity. CMB observation restricts this parameter to a maximum value of 1 \cite{Planck:2018constraitn}. The monodromy type potential goes back to the axion monodromy model if the parameters $n$ and $q$ are 1.

In the following sections, we use the effective masses in the analysis for convenience. The effective masses of the three models are defined as
\begin{equation}\label{eq:effective mass}
m^{2} = 
\begin{cases}
2n\Lambda^{2}\left(\frac{\Lambda}{M}\right)^{2}\left(\frac{\phi_{0}}{M}\right)^{2\left(n-1\right)}& \text{T model} \\
2^{2n+1}n\Lambda^{2}\left(\frac{\Lambda}{M}\right)^{2}\left(\frac{\phi_0}{M}\right)^{2\left(n-1\right)} & \text{E model} \\
q\Lambda^{2}\left(\frac{\Lambda}{M}\right)^{2}\left(\frac{\phi_{0}}{M}\right)^{2\left(n-1\right)}& \text{monodromy model}
\end{cases}
\end{equation}
which is the value of $\partial_{\phi_{0}}V/\phi_{0}$ when $\phi_{0} \ll M$ \cite{Lozanov:2018hjm}. Here we assume $\phi_{0}$ represents the inflaton field at the ending time of slow roll inflation, as well as the starting time of preheating.

\section{Floquet analysis}\label{Sec:Floquet}

We assume a homogeneous background $\bar{\phi}\left(t\right)$ and a small perturbation $\delta\phi\left(x,t\right)$ when the inflaton field oscillates at the bottom of the potential after inflation.
The expansion of the universe is neglected in Floquet analysis since the oscillating frequency $\omega\gg H$ 
\cite{Lin:2023inflation,Lozanov:2019review}. This implies that the period of oscillation is much shorter than the characteristic time scale of the expansion of universe. 

In linear analysis, the fluctuations of each Fourier mode are independent and can be easily solved numerically \cite{Amin:2014review,floquet-bands,Antusch:2017flz}. The equation of motion for the perturbative field is
\begin{equation}\label{eq:floquet}      
  \partial_{t}^{2} \delta\phi_{k}+\left[k^{2}+\partial_{\bar{\phi}}^{2}V\left(\bar{\phi}\right)\right]\delta\phi_{k}=0,
\end{equation}
which is known as the Mathieu equation (or Hill equation), characterized by its harmonic damping term. According to Floquet's theorem, the solution of the equation is
\begin{equation}
  \delta\phi_{k}=\mathcal{P}_{k+}\left(t\right)\exp\left(\mu_{k}t\right)+\mathcal{P}_{k-}\left(t\right)\exp\left(-\mu_{k}t\right)
\end{equation}
where $\mathcal{P}_{k\pm}$ is a periodic function determined by the initial conditions. The Floquet exponent $\mu_{k}$ is a complex number that only depends on the wavenumber $k$. If the real part of the exponent  $\Re_{\mu_{k}}\neq0$,  the solution is  unstable and exponentially growing, implying that a large number of resonant particles are produced.

\begin{figure*}[!htbp]
    \centering
    \subfigure{
    \begin{minipage}[b]{.3\textwidth}
    \centering
    \includegraphics[width=\linewidth, height=.8\linewidth]{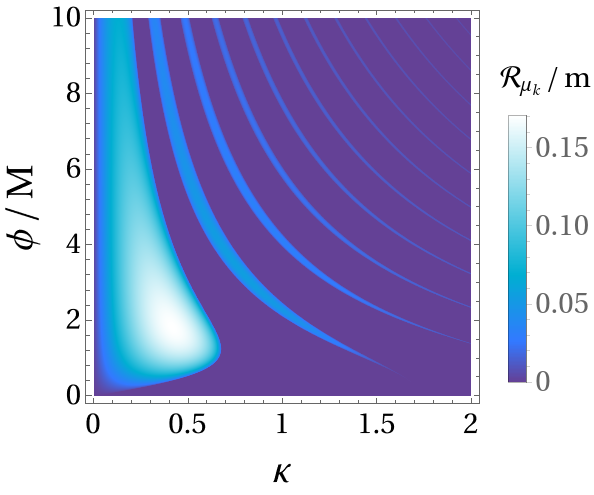}
    \caption*{n=1, q=0.25}
    \end{minipage}
    }
    \centering
    \subfigure{
    \begin{minipage}[b]{.3\textwidth}
    \centering
    \includegraphics[width=\linewidth, height=.8\linewidth]{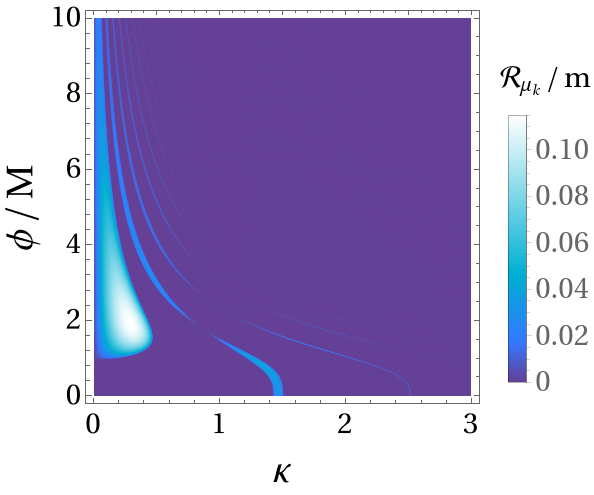}
    \caption*{n=1.5, q=0.25}
    \end{minipage}
    }
    \centering
    \subfigure{
    \begin{minipage}[b]{.3\textwidth}
    \centering
    \includegraphics[width=\linewidth, height=.8\linewidth]{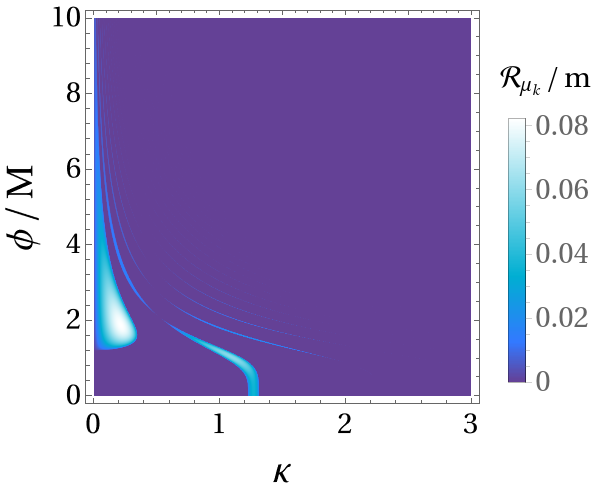}
    \caption*{n=2, q=0.25}
    \end{minipage}
    }
    \centering
    \subfigure{
    \begin{minipage}[b]{.3\textwidth}
    \centering
    \includegraphics[width=\linewidth, height=.8\linewidth]{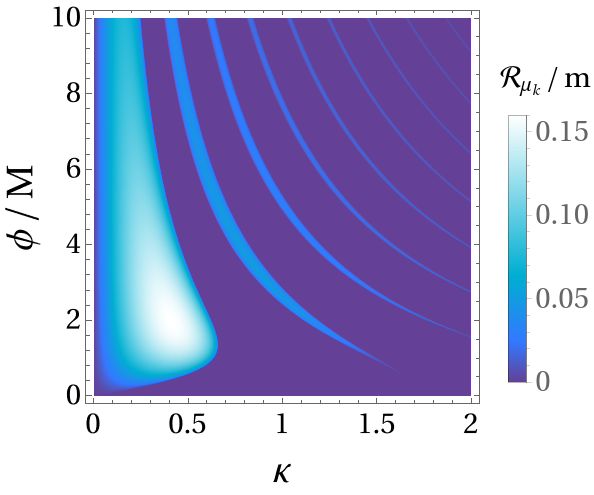}
    \caption*{n=1, q=0.5}
    \end{minipage}
    }
    \centering
    \subfigure{
    \begin{minipage}[b]{.3\textwidth}
    \centering
    \includegraphics[width=\linewidth, height=.8\linewidth]{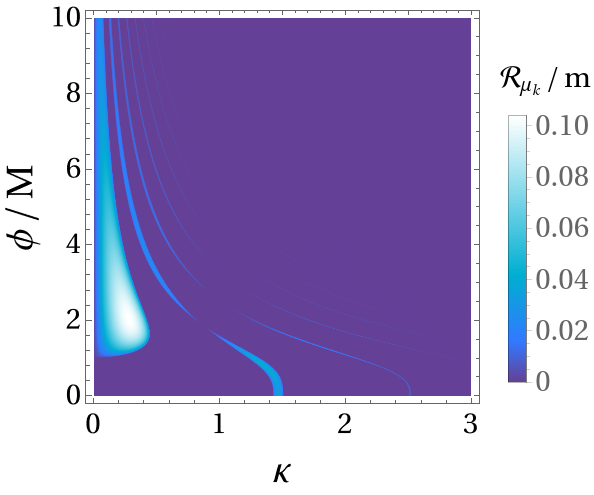}
    \caption*{n=1.5, q=0.5}
    \end{minipage}
    }
    \centering
    \subfigure{
    \begin{minipage}[b]{.3\textwidth}
    \centering
    \includegraphics[width=\linewidth, height=.8\linewidth]{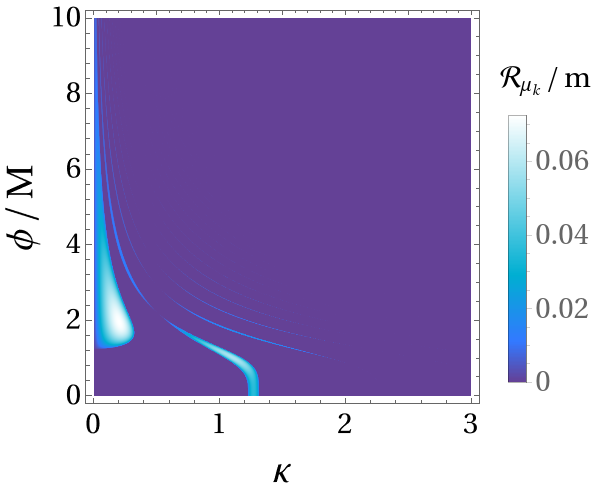}
    \caption*{n=2, q=0.5}
    \end{minipage}
    }
    \centering
    \subfigure{
    \begin{minipage}[b]{.3\textwidth}
    \centering
    \includegraphics[width=\linewidth, height=.8\linewidth]{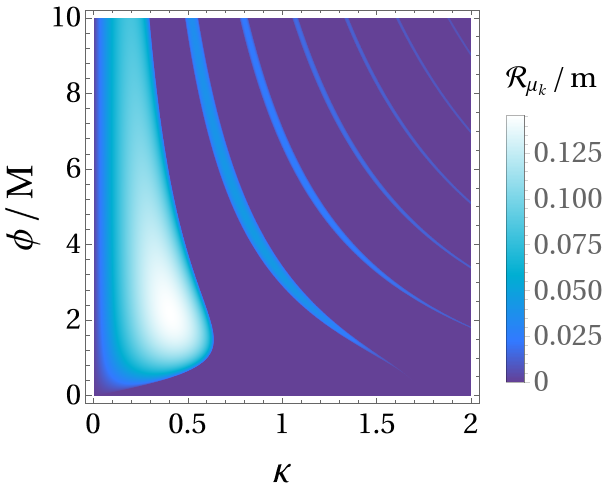}
    \caption*{n=1, q=0.75}
    \end{minipage}
    }
    \centering
    \subfigure{
    \begin{minipage}[b]{.3\textwidth}
    \centering
    \includegraphics[width=\linewidth, height=.8\linewidth]{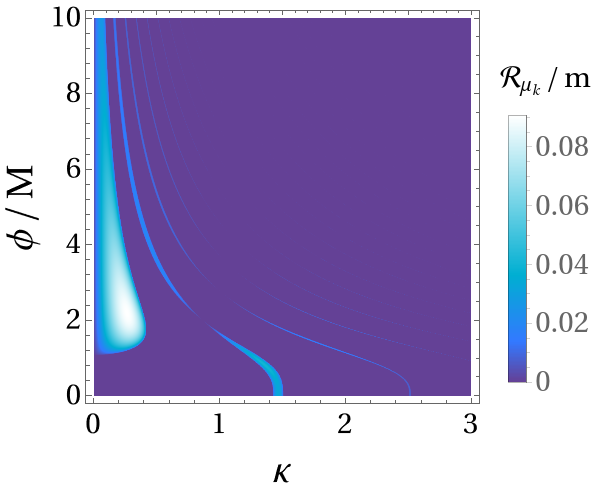}
    \caption*{n=1.5, q=0.75}
    \end{minipage}
    }
    \centering
    \subfigure{
    \begin{minipage}[b]{.3\textwidth}
    \centering
    \includegraphics[width=\linewidth, height=.8\linewidth]{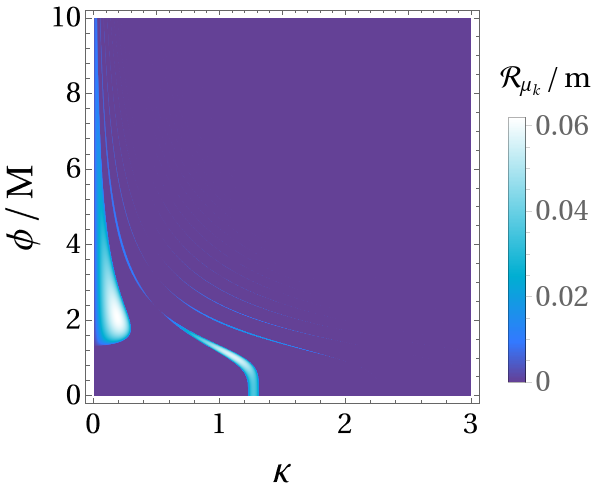}
    \caption*{n=2, q=0.75}
    \end{minipage}
    }
    \caption{The instability regions and Floquet exponents for monodromy model with $n=(1,1.5,2)$ and $q=(0.25,0.5,0.75)$. The longitudinal axis represents the amplitude of the oscillation of the inflaton field, while the horizontal axis represents the dimensionless wave number $\kappa=k/m$.} 
    \label{fig:floquet band}
\end{figure*}

Here we take the monodromy model as an example, and perform Floquet analysis with different model parameters.  The results of Floquet analysis are shown in Fig.~\ref{fig:floquet band}.  
The bright regions indicate the presence of unstable solutions. Due to their continuous regional distribution, they are called the instability bands. The instability bands shown in Fig.~\ref{fig:floquet band} are divided into a broad band and several narrow bands. 

The wide region in the Floquet plot in the Fig.~\ref{fig:floquet band} near the $k=0$ axis is the broad resonance band. The broad resonance band appears in all the parameter combinations we have selected, which 
represents the main region of the nonperturbative decay channel of inflaton. Compared with other unstable regions, the resonance strength of it is higher and the range is wider. The strong self-interaction with the inflaton background leads to a violation of the adiabatic conditions $\dot{\omega}/\omega^{2}\ge 1$ during each oscillation, resulting in the production of resonant particles in bursts within this band.

There are several narrow resonance bands away from the $k=0$ axis in all cases we have shown. 
For the $n=1$ cases, as shown in the first column in Fig. \ref{fig:floquet band}, the narrow resonance bands represent the other main nonperturbative decay channels in addition to the broad resonance band.
However in the case of $n>1$, as the instability zone moves, a series of new narrow resonance bands appear in the region $\phi \sim M$. Their resonance strength is lower and gradually disappears as the wave number $k$ increases. 
Since preheating occurs after the end of the slow roll inflation, with the amplitude of the oscillations $\phi \sim M$, such narrow bands, especially the first narrow band, dominate the evolution of the preheating. 

The parameter $n$ has a significant influence on the region of resonance bands. As the increase of $n$, the region of broad resonance band becomes smaller and some narrow resonance bands even disappear. The parameter $n$ has a weak influence on the resonance strength, causing it to decrease slightly with increasing $n$. As for the parameter $q$, it does not affect the region of resonance bands and has a minimal impact on the resonance strength. 
Therefore, in monodromy models, the shape of potential bottom, rather than that of the plateau, significantly influences the resonance bands. This conclusion is consistent with the Floquet analysis for $\alpha$-attractor potentials \cite{Lozanov:2018hjm}, of which the plateaus have the same shape but the bottom are affected by the parameter $n$, as shown in Fig. \ref{fig:potential}. We expect this statement could be generalized to other self-resonance models.

\section{Lattice Simulation}\label{Sec:Lattice method}
We perform a 3-dimensional (3D) lattice simulation of nonlinear real scalar field dynamics in an expanding universe. The public code \textbf{GABE}\footnote{https://cosmo.kenyon.edu/gabe.html} \cite{gabe} is used in our work.
In the simulation, the dynamics of inhomogeneous inflaton field is evolved by Eq. (\ref{eq:motion}), and the expansion of the universe is described by
\begin{equation}\label{eq:simulation motion}
\begin{gathered}
    H^{2}=\frac{1}{3m_{\rm pl}^{2}}\left<\frac{\dot{\phi}^{2}}{2}+\frac{\left(\nabla\phi\right)^{2}}{2a^{2}}+V\left(\phi\right)\right>_{s}.   
\end{gathered}
\end{equation}
Here the bracket denotes spatial average over the lattice. 

In this work, we focus on the dynamical evolution of the field and the non-adiabatic particles generation at the subhorizon scales.  The starting time of the simulation is set around the end of the inflation where the slow-roll approximation is broken.
We use a $N=256^3$ box in the lattice simulations and the lattice size $L=0.3/H$, which is always smaller than the hubble horizon scale. 

The nonlinear growth of the perturbation leads to inhomogeneity in the spatial distribution of the field, which is manifested in the enhanced resonance of the gradient term. We calculate the spatial distribution of energy density and transfer between different components including kinetic energy, potential energy, and gradient terms of the total energy density
\begin{equation}
    \rho=\frac{\dot{\phi}^{2}}{2}+\frac{\left(\nabla\phi\right)^{2}}{2a^{2}}+V\left(\phi\right).
\end{equation}

Another way to describe the evolutionary process is the equation of state of the universe. We use the spatially averaged equation of state
\begin{equation}\label{eq:EOS}
    w\equiv \frac{\left<p\right>_{s}}{\left<\rho\right>_{s}}=\frac{\left<\dot{\phi}^{2}/2-\left(\nabla\phi\right)^{2}/6a^{2}-V\right>_{s}}{\left<\dot{\phi}^{2}/2+\left(\nabla\phi\right)^{2}/2a^{2}+V\right>_{s}},
\end{equation}
where $\rho$ and $p$ represent the energy density and pressure of the inflaton field, respectively. 

In the following discussion, we will fix the value of parameter $M$ in these three models and consider different parameters $n$ or parameter combinations $\left[n,q\right]$ to simulate and analyze the results.

\section{Simulation of $\alpha$-attractor model}\label{sec:alpha attractor model}
The $\alpha$-attractor model has a power-law region near the origin ($\phi\ll M$) and a flat plateau far away from the bottom where the preheating process happens. In this section, we show the results of lattice simulation for  $\alpha$-attractor model.

\subsection{oscillons ($n=1$)}
For both T and E models from the $\alpha$ attractor, in the case of parameter $n=1$, the potential has a quadratic form near the origin and a region of platform when $\phi\gg M$, see Fig.~\ref{fig:potential}. The parameters in the simulation are listed in the first two rows of Table.~\ref{tab:alpha attractor models}.

\begin{table}[!htbp]
\begin{ruledtabular}
\begin{tabular}{ccccc}
\textrm{model}&
\textrm{n}&
\textrm{M$\left(m_{\rm pl}\right)$}&
\textrm{$\phi_{0}\left(m_{\rm pl}\right)$}\\
\colrule
$\alpha-T$ model & 1 & 0.01 & 0.04\\
$\alpha-E$ model & 1 & 0.01 & 0.034\\
$\alpha-T$ model & 1.5 & 0.01 & 0.038\\
$\alpha-E$ model & 1.5 & 0.01 & 0.035\\
\end{tabular}
\end{ruledtabular}
\caption{\label{tab:alpha attractor models} The parameters used in the lattice simulation. $\phi_{0}$ is the initial values of the background inflaton filed. } 
\end{table}

\subsubsection{back-reaction and structures formation}
During the preheating, the perturbations of the inflaton field are amplified exponentially. The resonance is sufficiently strong to induce back-reaction on the inflaton condensate, leading to the fragmentation. The nonlinear structures are generated in this process. 

\begin{figure}[!htbp]
    \centering
    \subfigure{
    \begin{minipage}[b]{.6\linewidth}
    \centering
    \includegraphics[width=\linewidth,height=\linewidth]{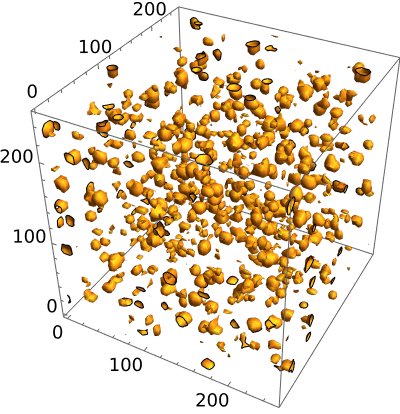}
    \caption*{(a) T model}
    \end{minipage}
    }
    \centering
    \subfigure{
    \begin{minipage}[b]{.6\linewidth}
    \centering
    \includegraphics[width=\linewidth,height=\linewidth]{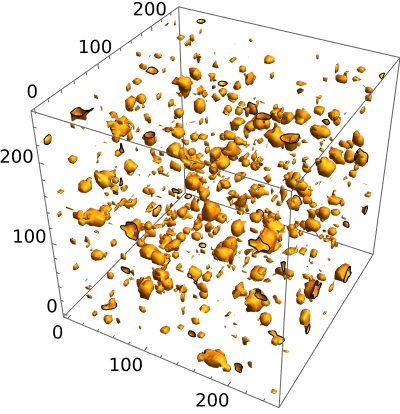}
    \caption*{(b) E model}
    \end{minipage}
    }
    \caption{Oscillons in lattice simulation of the $\alpha$-attractor T model and E model with parameters $n=1$. The figures show the energy density distribution when oscillons are fully formed. The yellow isosurface corresponds to the regions with 5 times the average energy density $\left<\rho\right>$. }
    \label{fig:alpha attractor model oscillons}
\end{figure}

As the inflaton condensates begin to fragment, numerous regions of energy concentration gradually emerge. These regions account for a large portion of the energy of the inflaton condensate and keep oscillating with the same frequency. 
Such structures, the so-called oscillons, exhibit a stable profile and survive millions of oscillations.
They usually form in a kind of potential models similar to what we chose with an expansion of a quadratic term near the origin and plateau near the large field value \cite{Amin:2012,Lozanov:2018hjm,a-efragment,Mahbub:2023faw,Alp-oscillon,Liu:2017huacospy.potential,Sang:2020kpd}. We can observe the oscillons in the simulation results through snapshot of the spatial distribution of energy density after the fully formation of oscillons, as shown in Fig.~\ref{fig:alpha attractor model oscillons}.

\subsubsection{equation of state and energy transfer}
\begin{figure*}[!htbp]
    \centering
    \subfigure{
    \begin{minipage}[b]{.4\textwidth}
    \centering
    \includegraphics[width=\linewidth]{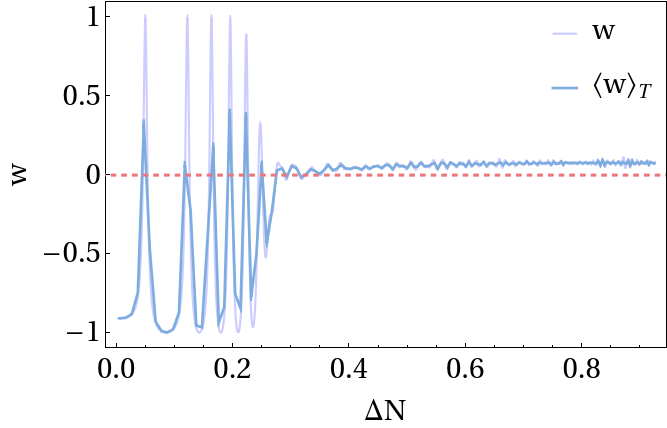}
    \caption*{(a) T model}
    \end{minipage}
    }
    \centering
    \subfigure{
    \begin{minipage}[b]{.4\textwidth}
    \centering
    \includegraphics[width=\linewidth]{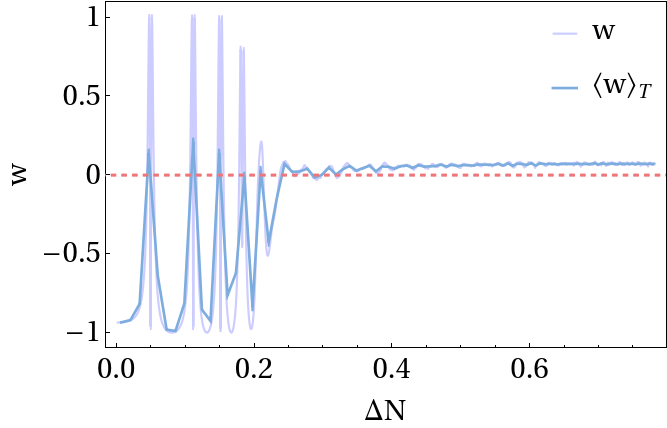}
    \caption*{(b) E model}
    \end{minipage}
    }
    \\
    \centering
    \subfigure{
    \begin{minipage}[b]{.4\textwidth}
    \centering
    \includegraphics[width=\linewidth]{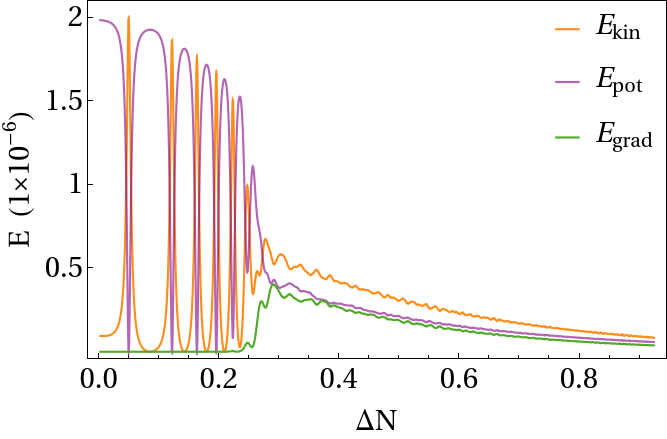}
    \caption*{(c) T model}
    \end{minipage}
    }
    \centering
    \subfigure{
    \begin{minipage}[b]{.4\textwidth}
    \centering
    \includegraphics[width=\linewidth]{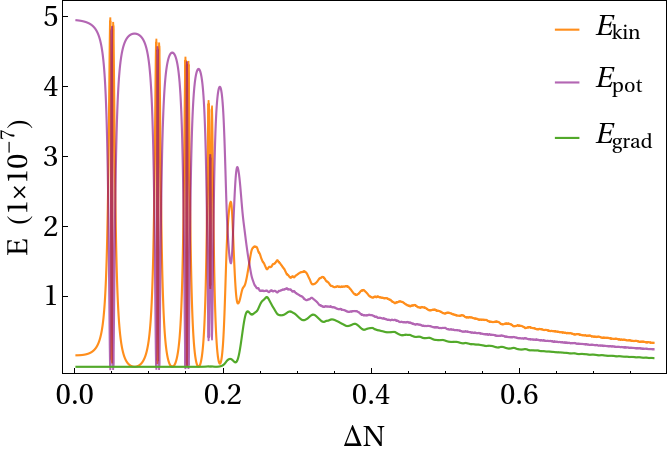}
    \caption*{(d) E model}
    \end{minipage}
    }
    \caption{\textbf{Top panel:} The evolution of the equation of state $w$ in $\alpha$-attractor T model (a) and E model (b) with parameters $n=1$. The purple curve $w$ is calculated by Eq. (\ref{eq:EOS}) from lattice. The blue curve $\left<w\right>_{T}$ is a time-average of $w$. \textbf{Bottom panel:} The evolution of energy in $\alpha$-attractor T model (c) and E model (d). The orange, purple and green curves are kinetic, potential and gradient terms, respectively.  }
    \label{fig:oscillon eos and energy transfer}
\end{figure*}

The equation of state $w$ is depicted by the blue curve in the top left and top right panels of Fig.~\ref{fig:oscillon eos and energy transfer} for T and E models, respectively. In both models, $w$ starts with a violent oscillation in the initial stage and then show a rapid damping, during which oscillons begin to emerge and form. The asymptotically approaching 0 of $w$ is corresponding to process that inflaton condensate fragment into oscillons. After the fully formation of oscillons, the curve of $w$ keeps slightly oscillating around $0$. Oscillons behave as pressureless dust with $w \sim 0$ and dominate the energy density of the universe for millions of oscillations.

Although most of the energy in the inflaton condensate flow into the oscillons, some remains stored in the in relativistic modes outside the oscillons. Therefore $w$ is approaching some value larger than 0 in Fig.~\ref{fig:oscillon eos and energy transfer}. The energy density stored in oscillons is evolving as $\propto \ a^{-3}$ and the energy stored in the relativistic modes decays as $a^{-4}$. If the two energy components are evolving independently, the energy of relativistic modes decreases more rapidly than that of oscillons, which should have led that $w$ decreases as $e^{-\Delta N}$. But the effect is not significant in  our simulations, and $w$ keeps nonzero $w$ when $\Delta N > 0.4$ in Fig.~\ref{fig:oscillon eos and energy transfer}. One possible reason is that we used a relatively short simulation time and we expect that the equation of state will finally goes to $w\to 0$ at a stage far outside the simulation time. 

\begin{figure}[!htbp]
    \centering
    \includegraphics[width=0.8\linewidth]{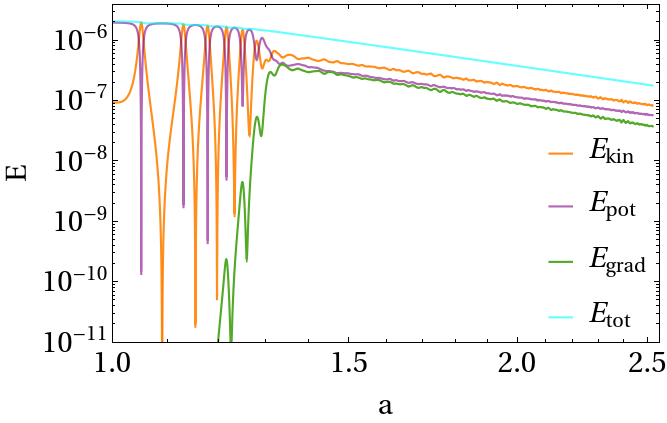}
    \caption{The evolution of energy with respect to scale factor in $\alpha$-attractor T model with parameters $n=1$. The orange, purple and green curves are kinetic, potential and gradient terms, respectively. The blue curve is total energy.}
    \label{fig:oscillon energy transfer loglog}
\end{figure}

The process is highly nonlinear and complex. The nonzero $w$ when $\Delta N > 0.4$ in Fig.~\ref{fig:oscillon eos and energy transfer} may not only simply result from the contributions of relativistic modes, but also it is possible that there are additional sources of relativistic modes. However, we do not expect the additional sources of relativistic modes are due to the decay of oscillons. Since oscillons are stable for a long time, the oscillon decay does not appear in our simulation. It can also be seen from the evolution of the energies. We show the time evolution of the energies with respect to scale factor in log-log scale. The total energy is depicted in blue curve. As shown in Fig.~\ref{fig:oscillon energy transfer loglog}, the energy follows a simple power-law scaling of the scale factor when $a > 1.5$, which indicates that the energy of oscillon and relativistic modes are independent of each other and there is no energy transfer from the oscillon to the relativistic modes.

The changing of the energy components is depicted in the bottom left and bottom right panels of Fig.~\ref{fig:oscillon eos and energy transfer} for T and E models, respectively. The orange, purple and green curves represent the kinetic energy, potential energy and gradient term of the inflaton field, respectively. The energy of each component decreases as the universe expands throughout the process. Once the fragmentation begins, the gradient term of the field is sharply amplified and reach its peak value, leading to a breakdown of system's adiabatic state \cite{Kasuya:2002adiabatic}. A large number of adiabatic particle oscillons are generated in every oscillation until the gradient term satisfies adiabatic conditions. Subsequently, these three energy components are continuously converted into each other and eventually reach a stable ratio when the oscillons are fully formed and stabilized.

While the condensate rapidly decays and breaks up, most of the energy component is transferred into oscillons. Simultaneously, a small amount of energy is redshifted into relativistic mode and gradually becomes negligible as a result of the expansion of universe. The above process occurs almost simultaneously.

\subsection{Transient ($n>1$)}
Next we will focus on the case of  parameter $n>1$. As depicted by three different colored curves for the three values of the parameters $n=1$, $1.5$, $2$ in the Fig.~\ref{fig:potential}, the potential shape in the bottom ($\phi \ll M$) are influenced by the parameters $n$.  As the parameter $n$ increases, the bottom region become flat and wide. The plateau regions ($\phi \gg M$) where the self-resonance happens and the slow roll inflation ends are not significantly changed by $n$. In this paper, we take $n=1.5$ as an example for $\alpha$-attractor T and E models, and show the results of lattice simulations, which has not been investigated in the previous literature. The parameters and initial conditions in the simulation are listed in the third and fourth rows of Table.~\ref{tab:alpha attractor models}.

\subsubsection{back-reaction and structures form}

Both T and E models exhibit intriguing phenomena in the simulations. The back-reaction on the inflaton condensates become apparent during the evolution. Some interconnected structures gradually formed, which is the same as the process of the back-reaction of oscillon cases ($n=1$). Subsequently the condensates fragment into another interesting nonlinear structure, which is called transient \cite{Lozanov:2018hjm,PhysRevD.2019}.

\begin{figure*}[!htpb]
    \centering
    \subfigure{
    \begin{minipage}[b]{.3\textwidth}
    \centering
    \includegraphics[width=0.9\textwidth,height=0.9\linewidth]{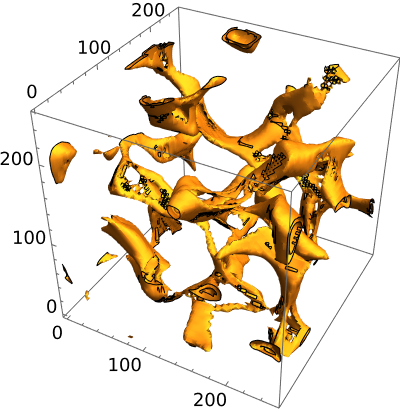}
    \caption*{a=1.28}
    \end{minipage}
    }
    \subfigure{
    \begin{minipage}[b]{.3\linewidth}
    \centering
    \includegraphics[width=0.9\textwidth,height=0.9\linewidth]{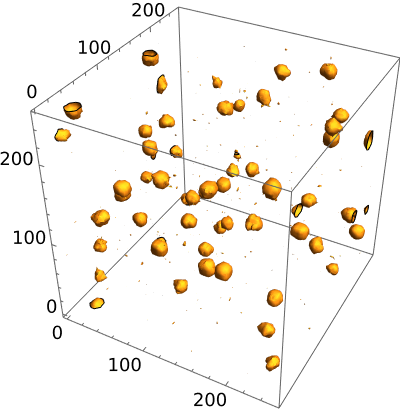}
    \caption*{a=2.72}
    \end{minipage}
    }
    \subfigure{
    \begin{minipage}[b]{.3\linewidth}
    \centering
    \includegraphics[width=0.9\textwidth,height=0.9\linewidth]{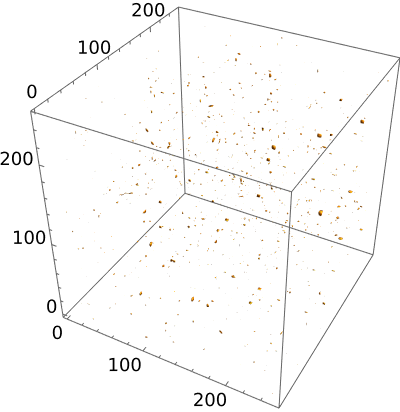}
    \caption*{a=3.91}
    \end{minipage}
    }
    \\
    \subfigure{
    \begin{minipage}[b]{.3\textwidth}
    \centering
    \includegraphics[width=0.9\textwidth,height=0.9\linewidth]{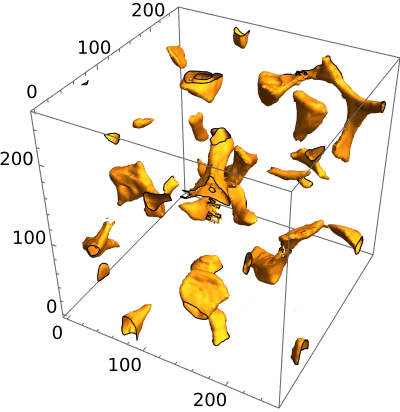}
    \caption*{a=1.26}
    \end{minipage}
    }
    \subfigure{
    \begin{minipage}[b]{.3\linewidth}
    \centering
    \includegraphics[width=0.9\textwidth,height=0.9\linewidth]{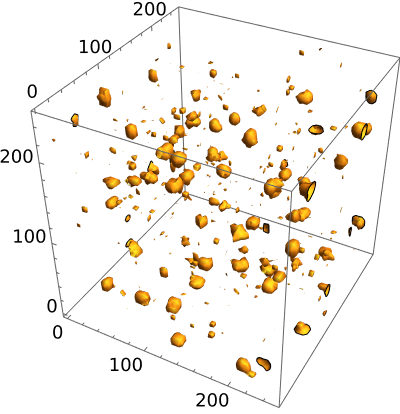}
    \caption*{a=3.23}
    \end{minipage}
    }
    \subfigure{
    \begin{minipage}[b]{.3\linewidth}
    \centering
    \includegraphics[width=0.9\textwidth,height=0.9\linewidth]{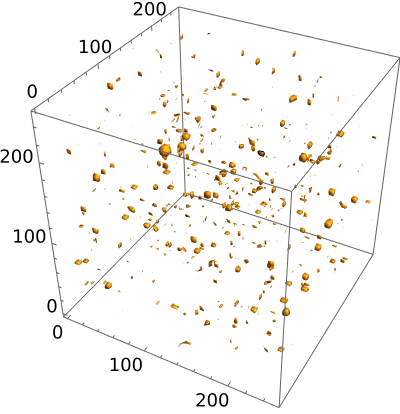}
    \caption*{a=4.98}
    \end{minipage}
    }
    \caption{Transients in lattice simulation of the $\alpha$-attractor T model (first row) and E model (second row) with parameters $n=1.5$ and $M=0.01m_{\rm pl}$. The figures show the energy density distribution before the formation, after the full formation and when the decay of transients. The yellow isosurface corresponds to the regions with 5 times the average energy density $\left<\rho\right>$. }
    \label{fig:transients with alpha attractor model}
\end{figure*}

As shown in Fig.~\ref{fig:transients with alpha attractor model}, transients form as soon as the fragmentation of the inflaton condensates begin. These structures exhibit approximately symmetric patterns and do not oscillate with time.  Throughout the process, their spatial positions remain nearly constant in the lattice. 
Initially, these regions of high energy density are extensive, but they subsequently decay and gradually transform into smaller individual structures. Ultimately, only a few remnants persist within the lattice.

\subsubsection{equation of state and energy transfer}

\begin{figure*}[!htbp]
    \centering
    \subfigure{
    \begin{minipage}[b]{.4\textwidth}
    \centering
    \includegraphics[width=\linewidth]{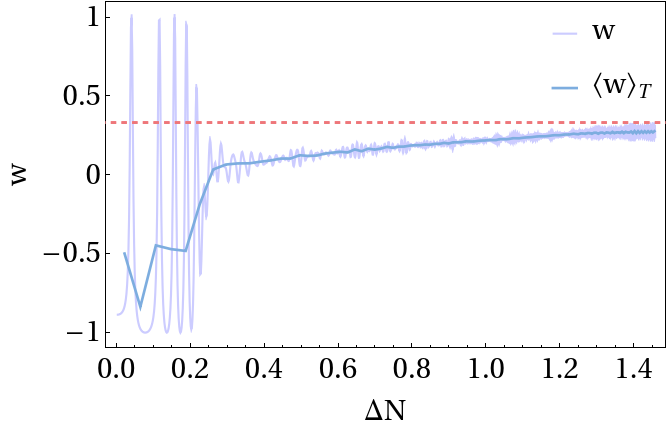}
    \caption*{(a) T model}
    \end{minipage}
    }
    \centering
    \subfigure{
    \begin{minipage}[b]{.4\textwidth}
    \centering
    \includegraphics[width=\linewidth]{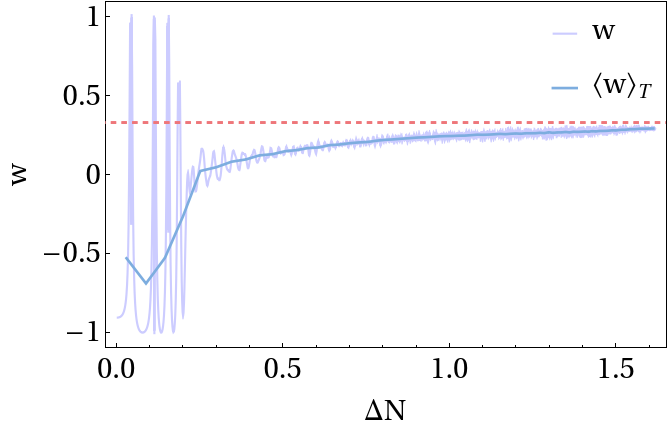}
    \caption*{(b) E model}
    \end{minipage}
    }
    \\
    \centering
    \subfigure{
    \begin{minipage}[b]{.4\textwidth}
    \centering
    \includegraphics[width=\linewidth]{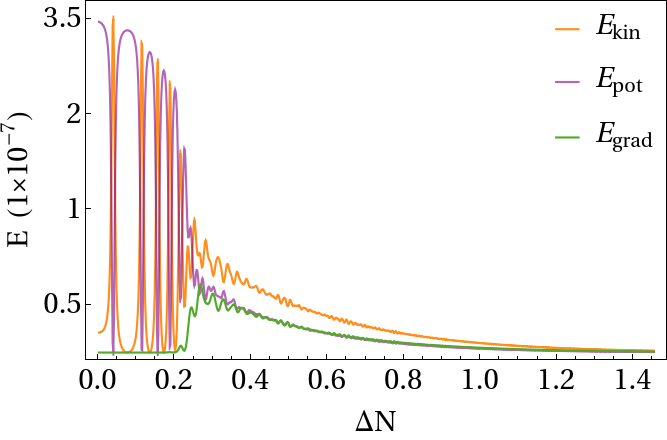}
    \caption*{(c) T model}
    \end{minipage}
    }
    \centering
    \subfigure{
    \begin{minipage}[b]{.4\textwidth}
    \centering
    \includegraphics[width=\linewidth]{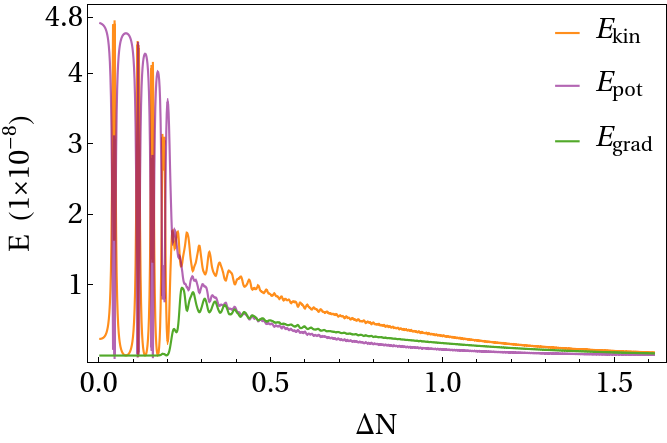}
    \caption*{(d) E model}
    \end{minipage}
    }
    \caption{\textbf{Top panel:} The evolution of the equation of state $w$ in $\alpha$-attractor T model (a) and E model (b) with parameters $n=1.5$. The purple curve $w$ is calculated by Eq (\ref{eq:EOS}) from lattice. The blue curve $\left<w\right>_{T}$ is a time-average of $w$. \textbf{Bottom panel:} The evolution of the  energy in $\alpha$-attractor T model (c) and E model (d). The orange, purple and green curves are kinetic, potential and gradient terms, respectively.  }
    \label{fig:alpha attractor models transient eos and energy transfer}
\end{figure*}

The equation of state $w$ is depicted in the top left and top right panels of Fig.~\ref{fig:alpha attractor models transient eos and energy transfer} for T and E models, respectively.  After the fragmentation of the condensates, $w$ oscillate around $0$ for a very brief moment. Subsequently $w$ begins to approach $1/3$ with the decay of transients. The entire process occurs within less than one $e$-fold of expansion.
The equation of state $w$ is approximated to that of radiation dominance in the process of transient decay and fragmentation. The equation of state $w$ in the transients case ($w \rightarrow 1/3$) is significantly different from the case of oscillons ($w \rightarrow 0$), as shown in Fig.~\ref{fig:oscillon eos and energy transfer}.

As depicted in the second row of Fig.~\ref{fig:alpha attractor models transient eos and energy transfer}, all of the energy stored declines as a result of the universe's expansion. The decrease in gradient term here occurs at a slower rate compared to other energy components and the gradient energy even surpasses potential energy after the moment of transient decay. 

From the simulation of the $\alpha$-attractor models, we find that the nonlinear dynamical behavior of the inflaton field during the preheating phase exhibits similarities in both the symmetric T model and the asymmetric E model.
The interesting nonlinear structures will be generated in both models, leading to the equation of state of the universe $w\to 0$ for parameter $n=1$ and $w\to 1/3$ for parameter $n>1$. 

It should be noted that the coherent oscillations gives $w\approx (n-1)/(n+1)$ for a potential of $V \propto \phi^{2n}$. Hence a potential with $n = 1.5$ leads to $w=0.2$. This conclusion is based on the assumption that the inflaton field is homogeneous and the gradient term vanishes. However during the preheating stage, inflaton field is inhomogeneous and the gradient contribution to the energy cannot be neglected. For the case of $n>1$, the energy density is stored in transients and relativistic modes. Since transients is shorter-lived, the energy stored in the relativistic modes will dominate after the decay of transients. Hence in the case of $n=1.5$, the equation of state will approach $w\to 1/3$.

\section{Simulation of monodromy models}\label{sec:axion monodromy models}

In this section, we will show the results of lattice simulation for monodromy model. The shape of the potential is shown in Fig.~\ref{fig:potential}. In the region near the bottom, the potential is given by
\begin{equation}
    V\left(\phi\right)=\Lambda^{4}\frac{q}{2n}\left|\frac{\phi}{M}\right|^{2n} \text{~if} \ \left|\phi\right|\ll M.
\end{equation} 
While in the region of $\phi\gg M$, the potential is approximated by 
\begin{equation*}
    V\left(\phi\right)=\Lambda^{4}\left|\frac{\phi}{M}\right|^{q} \  \text{~if} \ \left|\phi\right| \gg M, 
\end{equation*}
The model in the small field region depends on both $n$ and $q$, depend only on $q$ in large field region. The values of parameters using in the simulation is listed in Table.~\ref{tab:axion monodromy models}.

\begin{table}[h]
\begin{ruledtabular}
\begin{tabular}{ccccc}
\textrm{model}&
\textrm{n}&
\textrm{q}&
\textrm{M$\left(m_{\rm pl}\right)$}&
\textrm{$\phi_{0}\left(m_{\rm pl}\right)$}\\
\colrule
monodromy model & 1 & 1& 0.01 & 0.6\\
monodromy model & 1.5 & 1 & 0.01 & 0.71\\
monodromy model & 1.5 & 0.75& 0.01 & 0.56\\
monodromy model & 1.5 & 0.5 & 0.01 & 0.42\\
monodromy model & 1.5 & 0.25 & 0.01 & 0.31\\
monodromy model & 2 & 0.5 & 0.01 & 0.42\\
\end{tabular}
\end{ruledtabular}
\caption{\label{tab:axion monodromy models} The parameters used in the lattice simulation for monodromy models. $\phi_{0}$ is the initial values of the background inflaton filed.  } 
\end{table}

\subsection{structures form}

\begin{figure}[!htbp]
    \centering
    \subfigure{
    \begin{minipage}[b]{.6\linewidth}
    \centering
    \includegraphics[width=\linewidth]{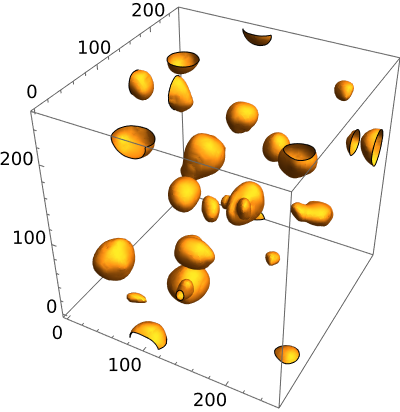}
    \caption*{(a) Oscillons of monodromy model}
    \end{minipage}
    }
    \subfigure{
    \begin{minipage}[b]{.6\linewidth}
    \centering
    \includegraphics[width=\linewidth]{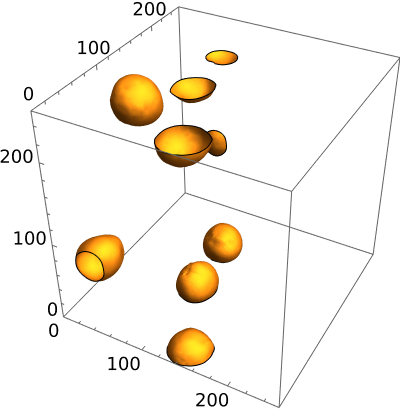}
    \caption*{(b) Transients of monodromy model}
    \end{minipage}
    }
    \caption{(a): Oscillons in lattice simulations of monodromy model with parameters $n=1$. (b): Transients in lattice simulations of monodromy model with parameters $n=1.5$. The figure shows the energy density distribution when nonlinear structures are fully formed. The yellow isosurface corresponds to the regions with 5 times the average energy density $\left<\rho\right>$. }
    \label{fig:monodromy model lattice result}
\end{figure}

Similar to the $n=1$ and $n>1$ cases in $\alpha$-attractor models, the parameter $n$ affects the formation of nonlinear structures in monodromy models. For example, if taking $q=1$, $V \propto \left|\phi\right|$ in the region of $\left|\phi\right| \gg M$ and $V \propto \left|\phi\right|^{2n}$ in the region of $\left|\phi\right| \ll M$. Fig.~\ref{fig:monodromy model lattice result} show the simulation of $n=1$ and $n=1.5$ cases in monodromy models, which are corresponding to the formation of oscillons and transients after the inflaton condensates fragmentation, respectively.

\subsection{effect of parameters $q$ and $n$}
We study the effect of parameters $q$ and $n$ on the nonlinear evolution by the simulations of different parameter values, see Table.~\ref{tab:axion monodromy models}.
To investigate the impact of parameter $q$, we fix parameter $n=1.5$ and run the simulation for $q=0.25$, $0.5$, $0.75$. 
As the parameter $q$ increases, the end time of slow roll inflation occurs at larger field values, see the $\phi_{0}$ values in Table.~\ref{tab:axion monodromy models}, which gives a larger initial amplitude of the oscillating sclar field. 
The back-reaction on inflaton condensates will occur at a later time. 
Therefore, as the parameter $q$ increases, the decay of inflaton field and subsequent nonlinear process are delayed.

\begin{figure*}[!htbp]
    \centering
    \subfigure{
    \begin{minipage}[b]{.3\textwidth}
    \centering
    \includegraphics[width=0.9\textwidth,height=0.9\linewidth]{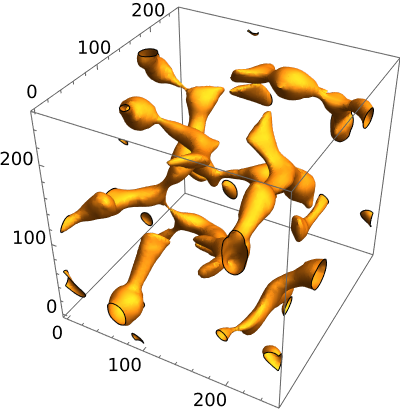}
    \caption*{a=3.73}
    \end{minipage}
    }
    \subfigure{
    \begin{minipage}[b]{.3\linewidth}
    \centering
    \includegraphics[width=0.9\textwidth,height=0.9\linewidth]{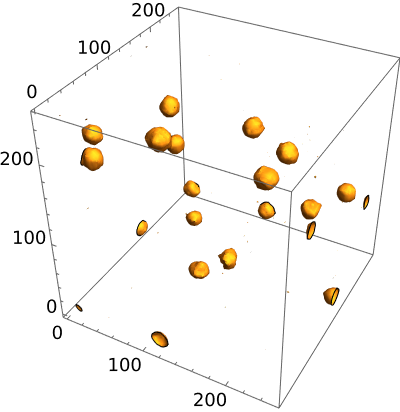}
    \caption*{a=8.73}
    \end{minipage}
    }
    \subfigure{
    \begin{minipage}[b]{.3\linewidth}
    \centering
    \includegraphics[width=0.9\textwidth,height=0.9\linewidth]{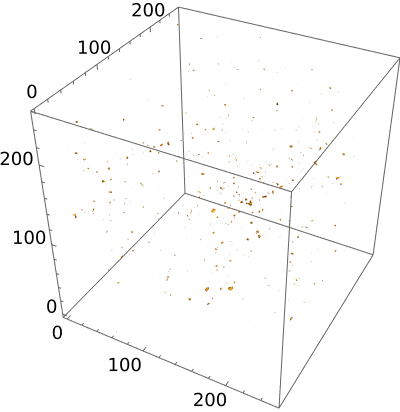}
    \caption*{a=10.73}
    \end{minipage}
    }
    \\
    \centering
    \subfigure{
    \begin{minipage}[b]{.3\textwidth}
    \centering
    \includegraphics[width=0.9\textwidth,height=0.9\linewidth]{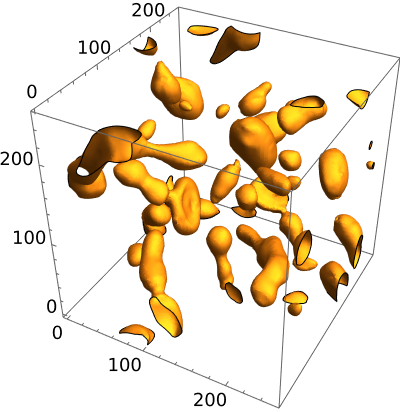}
    \caption*{a=4.89}
    \end{minipage}
    }
    \subfigure{
    \begin{minipage}[b]{.3\linewidth}
    \centering
    \includegraphics[width=0.9\textwidth,height=0.9\linewidth]{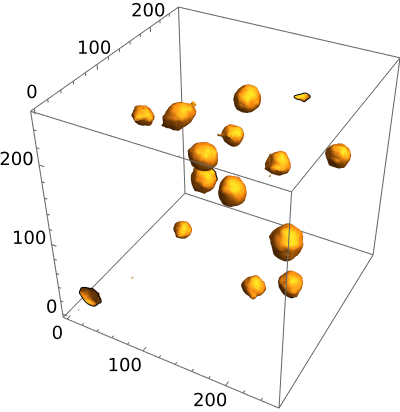}
    \caption*{a=9.58}
    \end{minipage}
    }
    \subfigure{
    \begin{minipage}[b]{.3\linewidth}
    \centering
    \includegraphics[width=0.9\textwidth,height=0.9\linewidth]{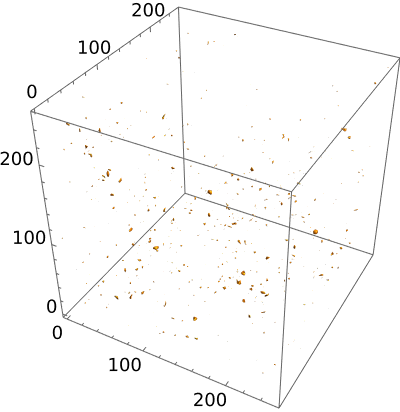}
    \caption*{a=14.34}
    \end{minipage}
    }
    \\
    \centering
    \subfigure{
    \begin{minipage}[b]{.3\textwidth}
    \centering
    \includegraphics[width=0.9\textwidth,height=0.9\linewidth]{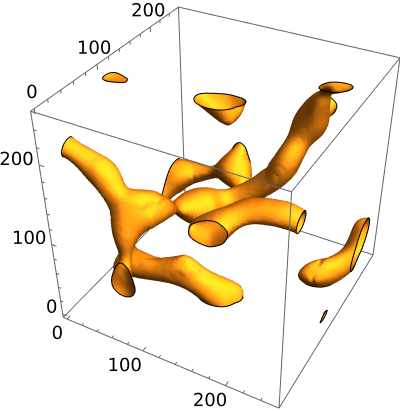}
    \caption*{a=6.28}
    \end{minipage}
    }
    \subfigure{
    \begin{minipage}[b]{.3\linewidth}
    \centering
    \includegraphics[width=0.9\textwidth,height=0.9\linewidth]{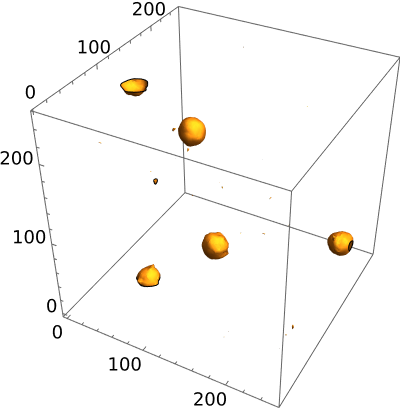}
    \caption*{a=14.34}
    \end{minipage}
    }
    \subfigure{
    \begin{minipage}[b]{.3\linewidth}
    \centering
    \includegraphics[width=0.9\textwidth,height=0.9\linewidth]{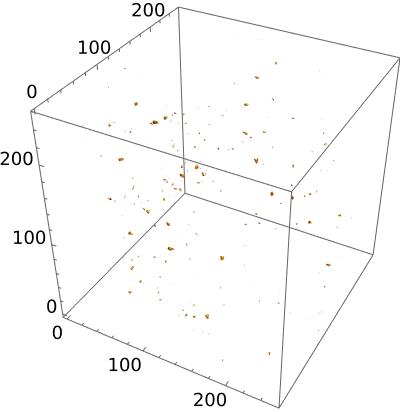}
    \caption*{a=18.18}
    \end{minipage}
    }
    \caption{Transients in lattice simulation of the monodromy model with parameters $n=1.5$ and $q=0.25$ (first row), $q=0.5$ (second row), $q=0.75$ (third row). The figures show the energy density distribution before the formation, after the full formation and when the decay of transients. The yellow isosurface corresponds to the regions with 5 times the average energy density $\left<\rho\right>$. }
    \label{fig:axion models transient n=1.5}
\end{figure*}

Fig.~\ref{fig:axion models transient n=1.5} shows the evolution of transients in the three cases of parameter $n=1.5$ and $q=0.25$, $0.5$, $0.75$.
The snapshot is taken at the time before the formation, after the full formation and when the decay of transients.
The lifetime of the transient can be defined as $\Delta N$ $e$-folds between the fragmentation of the inflaton condensates and the completely decay of the transient.
In the simulation, we find that the lifetime of transients are roughly equal for different $q$, around one $e$-fold.
From the scale factors given in Fig.~\ref{fig:axion models transient n=1.5}, the lifetime of transients are $\Delta N \sim 1.02, 1.07, 1.06$ for $q = 0.25, 0.5, 0.75$, respectively.
Hence, the parameter $q$ has limited influence on the lifetime of the transient.

\begin{figure*}[!htbp]
    \centering
    \subfigure{
    \begin{minipage}[b]{.3\linewidth}
    \centering
    \includegraphics[width=\linewidth]{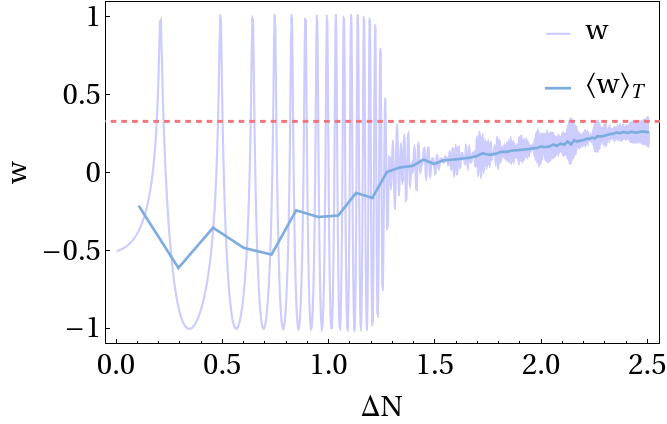}
    \end{minipage}
    }
    \subfigure{
    \begin{minipage}[b]{.3\linewidth}
    \centering
    \includegraphics[width=\linewidth]{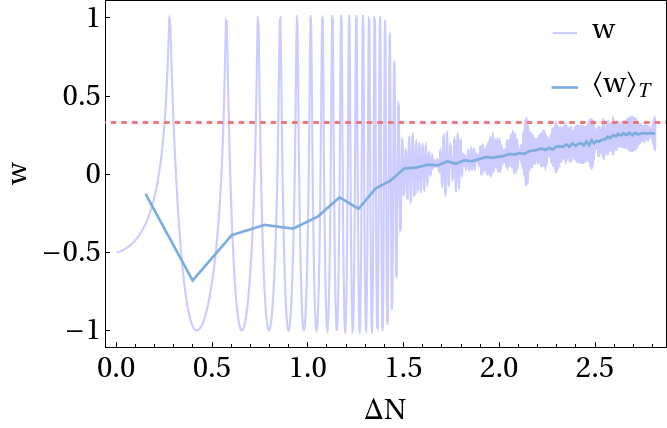}
    \end{minipage}
    }
    \subfigure{
    \begin{minipage}[b]{.3\linewidth}
    \centering
    \includegraphics[width=\linewidth]{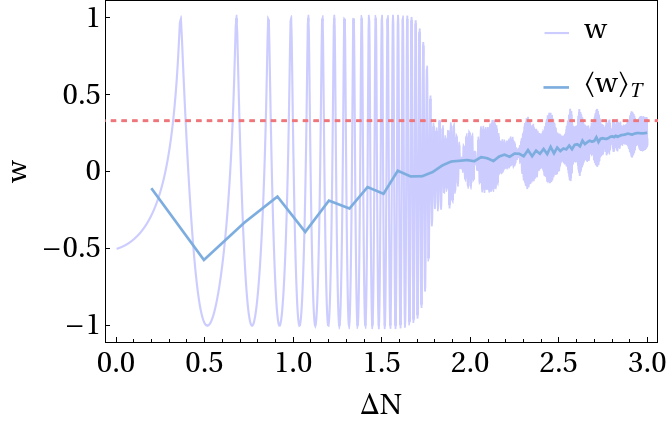}
    \end{minipage}
    }
    \subfigure{
    \begin{minipage}[b]{.3\linewidth}
    \centering
    \includegraphics[width=\linewidth]{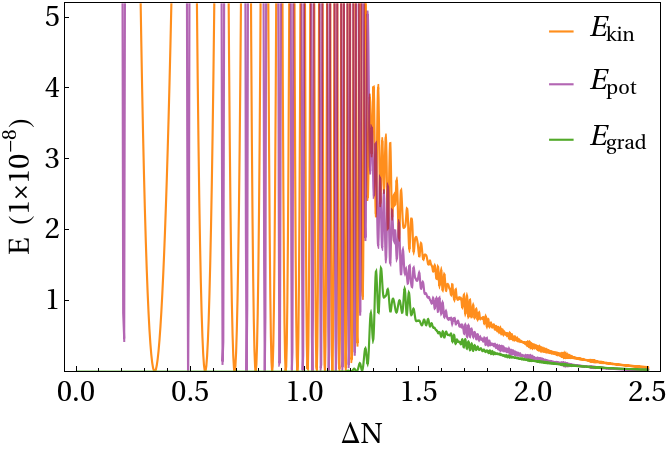}
    \caption*{(a) q=0.25}
    \end{minipage}
    }
    \subfigure{
    \begin{minipage}[b]{.3\linewidth}
    \centering
    \includegraphics[width=\linewidth]{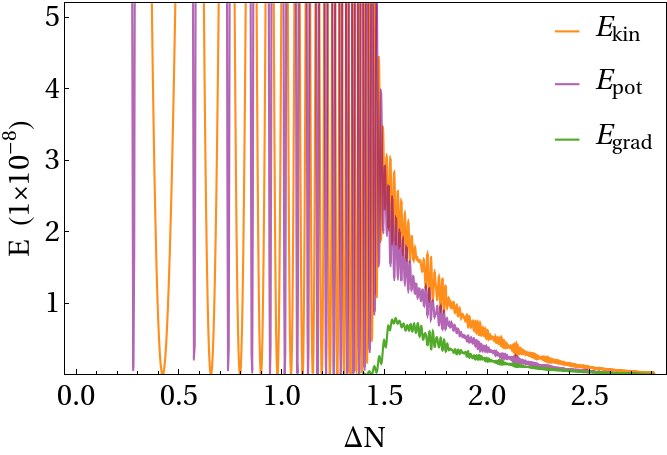}
    \caption*{(b) q=0.5}
    \end{minipage}
    }
    \subfigure{
    \begin{minipage}[b]{.3\linewidth}
    \centering
    \includegraphics[width=\linewidth]{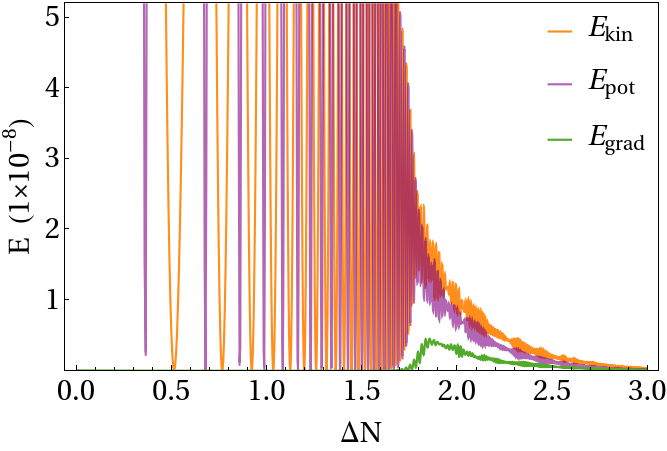}
    \caption*{(c) q=0.75}
    \end{minipage}
    }
    \caption{\textbf{Top panel:}The evolution of the equation of state $w$ in lattice simulation of the monodromy model ($n=1.5$) with different parameters $q=0.25$ (a), $q=0.5$ (b), $q=0.75$ (c). The purple curve $w$ is calculated by Eq (\ref{eq:EOS}) from lattice. The blue curve $\left<w\right>_{T}$ is a time-average of $w$. \textbf{Bottom panel:} The evolution of the energy in monodromy models with different parameters. The orange, purple and green curves are kinetic, potential and gradient terms, respectively. }
    \label{fig:axion transient eos and energy transefer n=1.5}
\end{figure*}

The first row in Fig.~\ref{fig:axion transient eos and energy transefer n=1.5} depicts the evolution of the equation of state for three different value of parameters $q$. 
In all three cases, the universe finally approaches to a radiation dominated state with the equations of state to be $1/3$. As the parameter $q$ decreases, the duration required to achieve a radiation-dominated universe is significantly reduced.
The second row in Fig.~\ref{fig:axion transient eos and energy transefer n=1.5} shows the energy components. As the parameter $q$ decreases, all of the energy components show a slight increase after the condensate fragment. 

\begin{figure*}[!htbp]
    \subfigure{
    \begin{minipage}[b]{.3\linewidth}
    \centering
    \includegraphics[width=0.9\textwidth,height=0.9\linewidth]{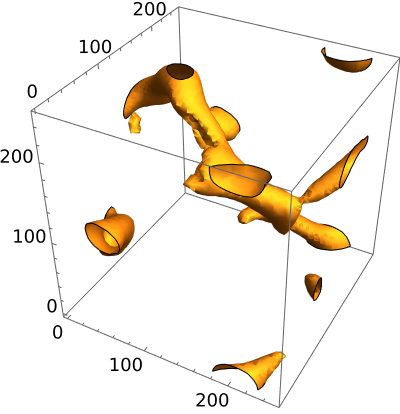}
    \caption*{a=4.24}
    \end{minipage}
    }
    \subfigure{
    \begin{minipage}[b]{.3\linewidth}
    \centering
    \includegraphics[width=0.9\textwidth,height=0.9\linewidth]{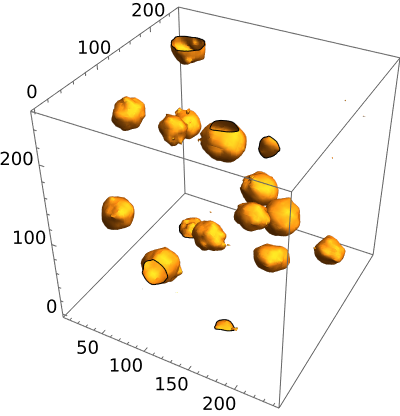}
    \caption*{a=6.80}
    \end{minipage}
    }
    \subfigure{
    \begin{minipage}[b]{.3\linewidth}
    \centering
    \includegraphics[width=0.9\textwidth,height=0.9\linewidth]{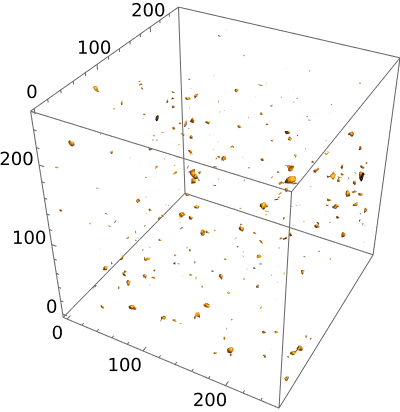}
    \caption*{a=8.64}
    \end{minipage}
    }
    \caption{Transients in lattice simulation of the monodromy model with parameters $n=2$ and $q=0.5$. The figures show the energy density distribution before the formation, after the full formation and when the decay of transients. The yellow isosurface corresponds to the regions with 5 times the average energy density $\left<\rho\right>$. } 
    \label{fig:axion models transient n=2}
\end{figure*}

Besides, we explore the effect of the parameter $n$ by comparing the simulations of parameters $n=1.5,q=0.5$ and parameters $n=2,q=0.5$. Fig.~\ref{fig:axion models transient n=2} shows the results of the simulation for parameters $n=2,q=0.5$.
Through the calculation, the lifetime of the transient in the case of parameter $n=2$ is $\Delta N \sim 0.71$, which is shorter than that in $n=1.5$ ($\Delta N \sim 1.07$). 
Therefore, the parameter $n$ has significant influence on the lifetime of the transient. With the parameter $n$ increasing, lifetime of the transient is largely reduced.

\begin{figure}[!htbp]
    \subfigure{
    \begin{minipage}[b]{.75\linewidth}
    \centering
    \includegraphics[width=\linewidth]{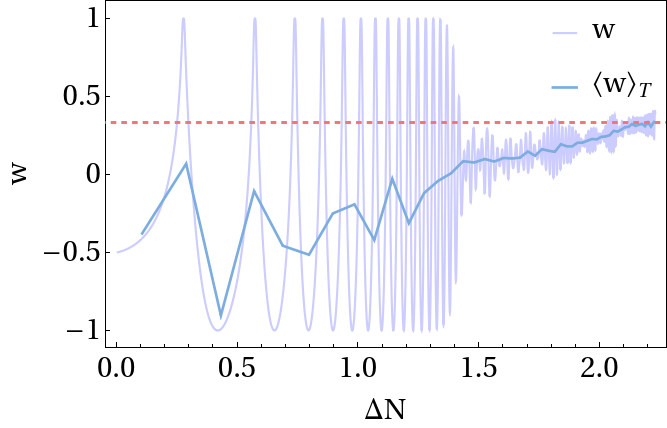}
    \end{minipage}
    }
    \subfigure{
    \begin{minipage}[b]{.75\linewidth}
    \centering
    \includegraphics[width=\linewidth]{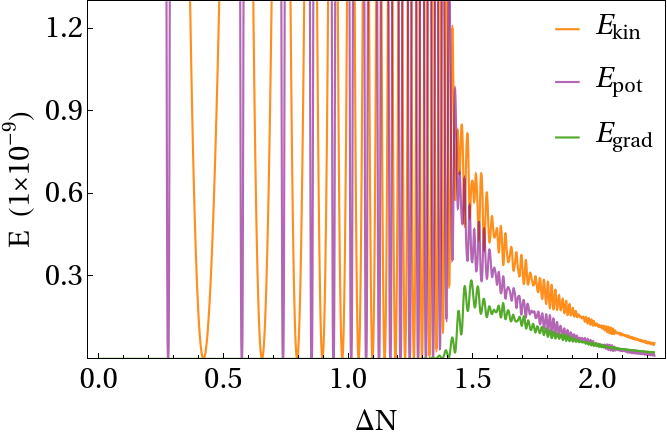}
    \end{minipage}
    }
    \caption{ \textbf{Top panel:} The evolution of the equation of state $w$ with parameter $n=2$ and $q=0.5$. The purple curve $w$ is calculated by Eq (\ref{eq:EOS}) from lattice. The blue curve $\left<w\right>_{T}$ is a time-average of $w$.  \textbf{Bottom panel:} The evolution of the energy in monodromy models. The orange, purple and green curves are kinetic, potential and gradient terms, respectively.} 
    \label{fig:axion transient eos and energy transefer n=2}
\end{figure}

As depicted in Fig.~\ref{fig:axion transient eos and energy transefer n=2}, the duration during which the equation of state $w$ approaches radiation dominance is shorter and the magnitudes of all energy components are relatively lower for $n=2$ than those $n=1.5$. This is consistent with the result of the Floquet analysis that an increase in the parameter $n$ will reduce the resonance slightly.

\section{Conclusion and discussion}\label{sec:conclusion}

In this paper, we investigated the nonlinear dynamics of the inflaton field during preheating in the $\alpha$-attractor T model, E model and monodromy model, which are favored by the observations. These models have a power-law form $\left|\phi\right|^{2n}$ near the origin ($\phi\ll M$) and a plateau in the region of $\phi\gg M$.
We presented the results of Floquet analysis in monodromy models and found that the potential parameters $n$ have a significant influence on the region of resonance bands, but have a weak influence on the resonance strength.  
Through the (3+1) dimensional lattice simulation, we performed a detailed analysis of the dynamic evolution and the nonlinear structure during preheating in all three models. In the $\alpha$-attractor models, symmetric potential in T model and the asymmetric potential in E model lead to similar nonlinear dynamics. We also find that with the increase of the potential parameter $n$ in monodromy model, the lifetime of the transient is largely reduced, but the potential parameter $q$ has limited influence on the lifetime of the transient.

With the different combinations of the parameters $q$ and $n$, we have investigated the influence of models parameters on the preheating. The parameters $q$ is only appear in the monodromy models. It mainly controls the region that $\phi\gg M$. In our simulation, we find that back-reaction process will be later with the parameters $q$ increases, but the lifetime of the transients we calculate in the simulations is almost unchanged. While the parameter $n$ appear in three models only controls the phase of preheating that inflaton condensate fragment into oscillon ($n=1$) or transients ($n>1$). As it increase, the lifetime of the transient from the fragmentation of the condensate to its decay of completely will be shorter. Compare to the parameters $q$, the effect of $n$ is more significant.

Note that such structures, oscillons and transients, can be generated under arbitrary parameter we selected. Therefore, we expect that such structures will exist in all of the models with a wider the parameter range. Besides, the phases of formation and decay of these structures can generate the gravitional waves, and whether difference combinations of parameter will influence their frequency or energy density remains an open question. We will stress these issues in the future work.

Our results reveal that the resonance structure, oscillon and transient, emerge following the decay of the inflaton condensate during preheating. These structures exhibit dust-like behavior characterized by an effective equation of state parameter $w\approx0$. The long-lived oscillon persists throughout our simulations, maintaining a matter-dominated universe $\left(w\approx 0\right)$ consistent with the current coherent oscillation paradigm. In contrast, transient decay within approximately one e-fold. This rapid decay of transients facilitates a gradual transition toward a radiation-dominated universe. In particular, in models with parameter $n>1$, the transient dynamics manifests a modified equation of state. This modification arises from nonlinear enhancement effects during the preheating phase, demonstrating the critical role of nonlinear interactions in the post-inflationary universe. Our analysis showed that the dynamics of the oscillons and transients is influenced mainly by the region near the minimum of the potential. Hence our conclusion should be generalized to a class of inflation models that exhibits a plateau at large field values $\left(\phi\gg M\right)$ and a power law form $\propto |\phi|^{2n}$  near small field values ($\phi\ll M$).

\section{Acknowledgement}
We would like to thank Shulan Li, Qingyang Wang and Shoupan Liu for helpful discussions. This work is supported by the National Natural Science Foundation of China (Grant Nos. 12175192, 12005184 and 12005183).

\newpage

\appendix

\nocite{*}

\bibliography{apssamp}

\end{document}